\documentclass[12pt]{article}

\usepackage[body={17.5cm, 22.5cm},right=2cm]{geometry}

\usepackage{float}

\usepackage{booktabs}
\usepackage{caption}

\usepackage{color}
\usepackage{graphicx}
\usepackage{epsf}
\usepackage{graphicx,epsfig}
\usepackage{amsmath}

\usepackage{multirow}

\usepackage{amsmath, amsthm, amssymb}

\usepackage{epsfig}
\usepackage{cite}
\usepackage{color,colordvi}

\def\K{K{\"a}hler}

\newcounter{hran}

\makeatletter
\renewcommand\section{\@startsection {section}{1}{\z@}%
                               {-3.5ex \@plus -1ex \@minus -.2ex}%
                               {2.3ex \@plus.2ex}%
                               {\normalfont\large\bfseries}}
\makeatother

\begin{document}\thispagestyle{empty}

\vspace{0.5cm}

\def\thefootnote{\arabic{footnote}}
\setcounter{footnote}{0}

\def\s{\sigma}
\def\nn{\nonumber}
\def\p{\partial}
\def\ls{\left[}
\def\rs{\right]}
\def\lc{\left\{}
\def\rc{\right\}}

\newcommand{\be}{\begin{eqnarray}}
\newcommand{\ee}{\end{eqnarray}}
\newcommand{\bi}{\begin{itemize}}
\newcommand{\ei}{\end{itemize}}
\renewcommand{\th}{\theta}
\newcommand{\bth}{\overline{\theta}}

\newcommand{\rf}[1]{(\ref{#1})}

\def\draftnote#1{{\color{red} #1}}
\def\bldraft#1{{\color{blue} #1}}

\begin{center}

\hskip 1cm

\vskip 2cm

{\LARGE \bf 
Is Imaginary Starobinsky Model Real?
 }
\\[1.5cm]
\vskip 1.5cm
{\normalsize  \bf    Renata Kallosh,  Andrei Linde,  Bert Vercnocke and Wissam Chemissany 
}
\\[1.1cm]

\vspace{.1cm}
{\small {  \it Stanford Institute for Theoretical Physics, and Department of Physics, Stanford University,\\ 382 Via Pueblo Mall, Stanford, CA 94305-4060, U.S.A.}}
\\

\end{center}

\vspace{1.2cm}

\begin{center}
{\small  \noindent \textbf{Abstract}} \\[0.5cm]

\end{center}
\noindent 
{\small
  We investigate the recently proposed possibility of chaotic inflation with respect to the imaginary part of the field $T$ in a supersymmetric embedding of the Starobinsky model.  We show that the stage of rapid expansion driven by ${\rm Im} \ T$ in this model ends almost instantly, and the subsequent stages of inflation are driven by the real part of the field $T$, as in the standard Starobinsky model. Thus, the  Starobinsky model and its supersymmetric generalizations remain disfavored by the recent BICEP2 data.

\def\thefootnote{\arabic{footnote}}
\setcounter{footnote}{0}

\baselineskip= 18pt
\newpage 
\section{Introduction}\pagenumbering{arabic}

During the last few years we witnessed several radically different trends in the development of inflationary cosmology. Prior to the Planck data release \cite{ade}, there were rumors of the possible discovery of large non-Gaussianity, which was supposed to rule out 99\% of all existing inflationary models. Many people worked on the remaining 1\% of the models, which where rather exotic and complicated. The Planck data revealed that the $f^{\rm local}_{\rm NL}$ is as small as one could expect from the simplest inflationary models, so we were able to return to the investigation of more regular versions of inflationary theory. However, the results by Planck and other experiments suggested that the amplitude of tensor modes is very small, $r \lesssim 0.1$. Consequently, attention was attracted to the models predicting very small values of $r$, such as the Starobinsky model, new inflation, Higgs inflation, and many versions of string theory inflation predicting $r \ll 0.01$. The simplest versions of the chaotic inflation scenario, such as the model with a quadratic potential ${m^{2}\over 2} \phi^{2}$ \cite{chaotic} were disfavored.

Now, in a surprising twist of fortune, the chaotic inflation models such as ${m^{2}\over 2} \phi^{2}$ are at the center of attention, 
but all models predicting small values of $r$ are strongly disfavored by the recent  BICEP2 data, which found B-modes and concluded that $\ensuremath{r = 0.2^{+ 0.07}_{- 0.05}}$,  \cite{bicep}. In particular, according to \cite{bicep}, the Starobinsky model \cite{Starobinsky:1980te,Starobinsky:1983zz}, predicting $r \sim 0.004$, is disfavored at a level greater than $5\sigma$.

It is too early to discard models with small $r$ which have been popular since the latest Planck data release, especially since there is a broad class of theories continuously interpolating between the models with $r < 0.01$ and the models with $r \gtrsim 0.2$ \cite{Kallosh:2013tua,Kallosh:2013yoa}. Also, some of the models predicting $r < 0.01$ can be modified in a way making them consistent with the BICEP2 data.  

In a recent paper \cite{Ferrara:2014ima} it was argued that the supersymmetric generalization of the theory $R+ R^{2}$ developed by Cecotti back in 1987 \cite{Cecotti:1987sa} may allow two different inflationary regimes. One of them, driven by the real part of the scalar field $T$ present in this model, coincides with Starobinsky inflation, predicting $r \sim 0.004$, whereas another one, driven by the imaginary part of the field $T$, describes chaotic inflation with the quadratic potential  \cite{chaotic}, predicting $r \sim 0.15$. 

Of course, the possibility of such a regime does not make the original Starobinsky model consistent with the BICEP2 results, but nevertheless it would be very interesting  if its supersymmetric generalization, containing 3 extra scalar degrees of freedom as compared to the original Starobinsky model, were able to describe an inflationary regime providing a good match to the recent BICEP2 data. 
In this paper we will analyze this issue and show that there is no inflationary regime driven by the imaginary part of the field $T$ because large field ${\rm Im} T$ destabilizes the real part of the field $T$ in a way to be described shortly. It is possible to solve this problem by stabilizing the real part of the field $T$, but this requires a strong modification of the model of \cite{Cecotti:1987sa}, after which it ceases to have any relation to the original Starobinsky model $R+R^{2}$, as well as to pure higher derivative supergravity.

\section{Starobinsky model and Cecotti's  pure higher derivative curvature model\cite{Cecotti:1987sa}}

The original version of the Starobinsky model was based on the investigation of Einstein gravity including the contribution from the conformal anomaly  \cite{Starobinsky:1980te}. The  goals of this model were in a certain sense opposite to the goals of inflationary cosmology. Instead of attempting to solve the homogeneity and isotropy problems, Starobinsky considered the model of  the universe which was homogeneous and isotropic from the very beginning, and emphasized that his scenario was ``the extreme opposite of Misner's initial chaos''   \cite{Starobinsky:1980te}. An additional issue was that his model was non-singular, so its solution could be continued to $t \to -\infty$. On the other hand, this solution was unstable  \cite{Starobinsky:1980te,Mukhanov:1981xt}, which made its existence at $t \to -\infty$ problematic.

Then, in 1983, this model was streamlined, its Lagrangian  was written in the form
\be\label{star}
L={ \sqrt{-g}} \left({1\over 2} R+{R^2\over 12M^2}\right) \, ,
\ee
where $M \ll 1$  \cite{Starobinsky:1983zz}, and the initial conditions for inflation in this model were formulated along the lines of the chaotic inflation scenario  \cite{chaotic}.

This theory is conformally equivalent to canonical gravity plus a scalar field $\phi$ \cite{Whitt:1984pd}.
Making the transformation $\tilde{g}_{\mu\nu} = (1 + \phi/3M^2) g_{\mu\nu}$
and the field redefinition $\varphi = \sqrt{\frac{3}{2}} \ln \left( 1+ \frac{\phi}{3 M^2} \right)$,
one finds the equivalent Lagrangian 
\be\label{whitt}
L=\sqrt{-\tilde{g}}\left[{1\over 2}\tilde{R} - {1\over 2}\partial_{\mu} \varphi\partial^{\mu} \varphi - \frac{3}{4} M^2 \left(1- e^{-\sqrt{2/3}\,\varphi}\right)^2 \right] \, .
\ee

In 1987, Cecotti proposed a supergravity embedding of the theory $R+R^{2}$  \cite{Cecotti:1987sa}.  The higher derivative model of \cite{Cecotti:1987sa}  depends on a chiral curvature superfield ${\cal R}$ and has no interaction with matter, it is therefore  a  unique pure higher curvature supergravity model with up to 4 derivatives. A year later a different  formulation of the supersymmetric version of the $R+R^{2}$ was given by Cecotti, Ferrara, Porrati and Sabharwal in \cite{Cecotti:1987qe} in the context of a new minimal formulation.

It was shown in \cite{Cecotti:1987sa} that upon a dual transformation the higher derivative model  depending only on a curvature superfield is  given by the standard 2-derivative supergravity interacting with two chiral matter multiplets, $T$ and $C$, see also more recent and detailed discussion of this in  \cite{CK}, \cite{fkvp}. This dual supergravity model is defined by the following K\"ahler potential and superpotential
\be
\label{K1}
K = -3\, \ln\Big{(}T + \overline{T} - \overline{C} C\Big{)}\ ,  \qquad W= 3 M C(T-1) \ .
\ee
The  cosmological  analysis of this supergravity model was performed in \cite{Kallosh:2013lkr} where it was shown that by adding quartic in $C$ terms to the K\"ahler potential it is possible to stabilize the model at $C=\bar C= {\rm Im} \ T=0$. The role of the inflaton is played by ${\rm Re} \ T$ and the bosonic model with $C=\bar C= {\rm Im} \ T=0$ coincides with the Starobinsky model. 

In \cite{Ferrara:2014ima} it was suggested to study a different regime of this supergravity model \cite{Cecotti:1987sa}, where $C=\bar C= {\rm Re} \ T-1=0$. The supersymmetric model of Cecotti \cite{Cecotti:1987sa} shown in eq. (\ref{K1}) at $C=\bar C= {\rm Re} \ T-1=0$ was called  in \cite{Ferrara:2014ima}, `The Imaginary Supersymmetric Starobinsky Model'. This name is somewhat misleading since it may suggest that the stage of inflation driven by the imaginary part of the field $T$ in this scenario is described by  the Starobinsky model; this is not the case.

Note also that  the higher derivative supergravity model \cite{Cecotti:1987qe}  has another dual version which was studied in the cosmological context recently in \cite{Farakos:2013cqa}, \cite{Ferrara:2013rsa}. It is a standard 2-derivative supergravity interacting with the vector multiplet. Only one real scalar field is available there and therefore it leads only to Starobinsky model with low level of $r$, there are no other scalars which could be used as inflatons to increase $r$.

\section{Inflation}

Following \cite{Ferrara:2014ima}, one can represent the bosonic part of the Cecotti model as follows:
\be
\label{OM71}
e^{-1}{\cal L} = \frac{1}{2} R  -  \frac{1}{2} \partial_{\mu} \phi  \partial^{\mu} \phi -\frac{1}{2} e^{-2\sqrt{2/3}\,  \phi}\partial_\mu b \partial^\mu b-\frac{1}{2}M^2 e^{-2\sqrt{2/3}\,  \phi}\,  b^2
-\ \frac{3}{4} M^2 \,  
\bigl(1-e^{-\sqrt{2/3}\,  \phi}\bigr)^2.
\ee
Here $T=e^{\sqrt{2/3}\, \phi} + i \sqrt{2/3} \, b$,  and  in our model $\phi=0$ is a minimum of the potential as our superpotential has a minimum at $T=1$.
Note that at $b=0$ we recover the Starobinsky model \cite{Starobinsky:1983zz}, \cite{Whitt:1984pd}.
 
In fact, eq. \rf{OM71} gives an incomplete description, because the model also includes the field $C$ which requires stabilization. It can be achieved along the lines of   \cite{Kallosh:2013lkr}. This will not be important for our main conclusions.

As one can see, the potential of this model with respect to the axion field $b$ is quadratic, so if it were possible to separate its evolution from the evolution of the field $\phi$, one would have a perfect realization of the simplest chaotic inflation model with the quadratic potential \cite{chaotic}, which would provide a good fit to the recent BICEP2 data.

At first glance, this could have happened in this theory because  the potential $\frac{3}{4} M^2\bigl(1-e^{-\sqrt{2/3}\,  \phi}\bigr)^2$ tends to keep the field $\phi$ captured at the minimum of this potential where $\phi=0$ \cite{Ferrara:2014ima}. However, chaotic inflation driven by the field $b$ requires initial condition $b^2\gg 1$. For $b^2\gg 1$, $\phi = 0$, the term $\frac{1}{2}M^2 e^{-2\sqrt{2/3}\,  \phi}\,  b^2$ in the scalar field potential  is much greater than the stabilizing term $\frac{3}{4} M^2\bigl(1-e^{-\sqrt{2/3}\,  \phi}\bigr)^2$, so the stabilization provided by this term does not work at that stage. The complicated nature of this potential is shown in Fig. \ref{fig1}.

Therefore, if one follows  \cite{Ferrara:2014ima} and assumes that initially $b^2\gg 1$ and $\phi =0$, one can ignore the stabilizing term   $\frac{3}{4} M^2\bigl(1-e^{-\sqrt{2/3}\,  \phi}\bigr)^2$ for the description of the initial stages of  the cosmological evolution. The term $\frac{1}{2} e^{-2\sqrt{2/3}\,  \phi}\partial_\mu b \partial^\mu b$ also induces a potential for the field $\phi$. One could expect that this term may compensate the influence of the term $\frac{1}{2}M^2 e^{-2\sqrt{2/3}\,  \phi}\,  b^2$,  but if inflation does occur, the term $\frac{1}{2} e^{-2\sqrt{2/3}\,  \phi}\partial_\mu b \partial^\mu b$  gives a subdominant contribution to the potential of the field $\phi$, simply because the kinetic and gradient terms of the field $b$ are much smaller than the potential energy of this field  during inflation. In the slow-roll approximation, one can also ignore the mixed terms proportional to $\dot\phi\dot b$. 

In order to study the evolution of the fields in this model, one should go beyond the slow-roll approximation. We will do it shortly. But it is easier to understand what happens if one attempts first to preserve as much as possible of the original assumptions of \cite{Ferrara:2014ima}, and then check whether these assumptions actually work. So let us study the evolution of the field $\phi$, but assume that this field changes slowly, which is a generalization of the assumption of \cite{Ferrara:2014ima} that the field $\phi$ does not move at all. Then one can use the slow-roll approximation for the investigation of the evolution of the field $\phi$.

In this case, the equation of motion for the scalar field $\phi$ during inflation at $b\gg 1$ (if this regime actually happens)  acquires a simple familiar form
\be
3H\dot\phi = - V_{,\phi} \ ,
\ee
where the potential is $V(\phi,b)  = \frac{1}{2}M^2 e^{-2\sqrt{2/3}\,  \phi}\,  b^2$, and $V_{,\phi} = -\sqrt{2/3}\, M^2 e^{-2\sqrt{2/3}\,  \phi}\,  b^2$. It is convenient to make the standard change of variables $H dt = dN$. Here $N$ is the number of e-foldings,  the positive sign corresponds here to the choice of the number of e-foldings $N$ growing from the beginning of inflation. Since $3H^{2} = V$, one can replace the equation for the field $\phi$ with the equation written in a more transparent form:
\be
{d\phi\over dN} = - {V_{,\phi}\over V} = 2\sqrt{2\over  3} \ .
\ee
This is the main result, which makes everything else obvious: During each e-folding, the field $\phi$ grows by  $ 2\sqrt{2/3}$. Therefore during $N$ e-foldings, the potential $V(\phi,b)  = \frac{1}{2}M^2 e^{-2\sqrt{2/3}\,  \phi}\,  b^2$ of the field $b$ drops exponentially, by a factor of $e^{-8N/3}$, in addition to the decrease due to the rolling of the field $b$ down towards $b = 0$. This immediately kills the stage of chaotic inflation driven by the field $b$. After that, the field $\phi$ stops growing linearly with $N$, it becomes captured by the potential $\frac{3}{4} M^2\bigl(1-e^{-\sqrt{2/3}\,  \phi}\bigr)^2$ and start rolling back to the minimum of this potential $\phi=0$. The remaining stage of inflation, which can be very long if the initial value of the field $b$ was very large, is described by inflation driven by the standard Starobinsky-Whitt potential $\frac{3}{4} M^2\bigl(1-e^{-\sqrt{2/3}\,  \phi}\bigr)^2$. Therefore such scenario leads to the same consequences as the Starobinsky model, i.e. to $r \sim 0.004$, which is disfavored by BICEP2.

\begin{figure}[h!t!]
\centering
\includegraphics[width=.48\textwidth]{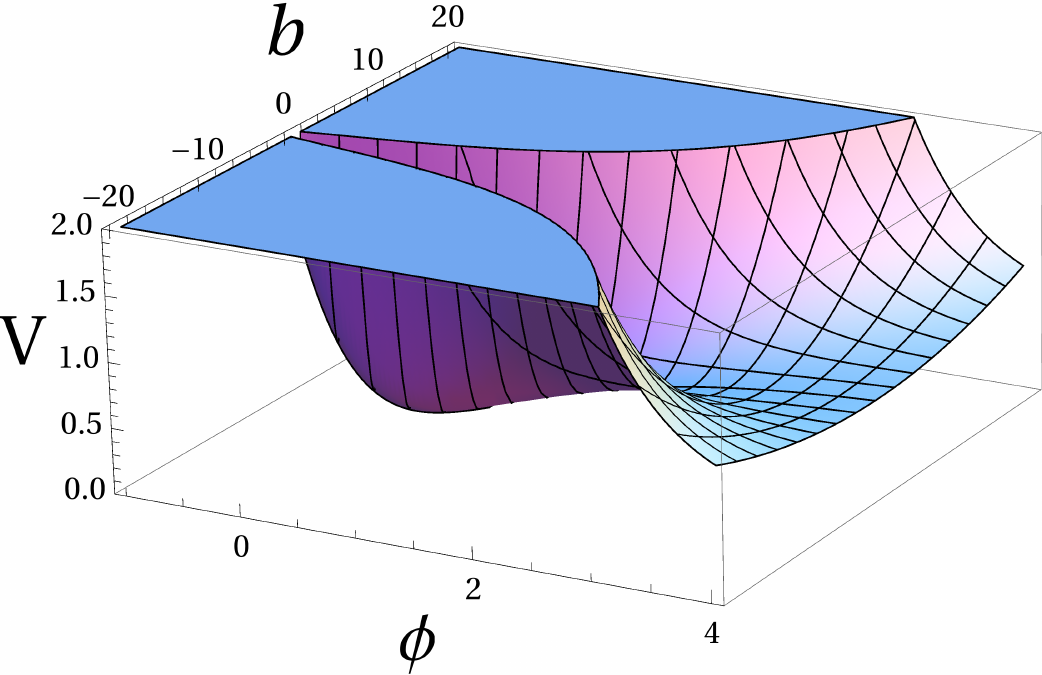}
\hspace{.05\textwidth}
\includegraphics[width=.45\textwidth]{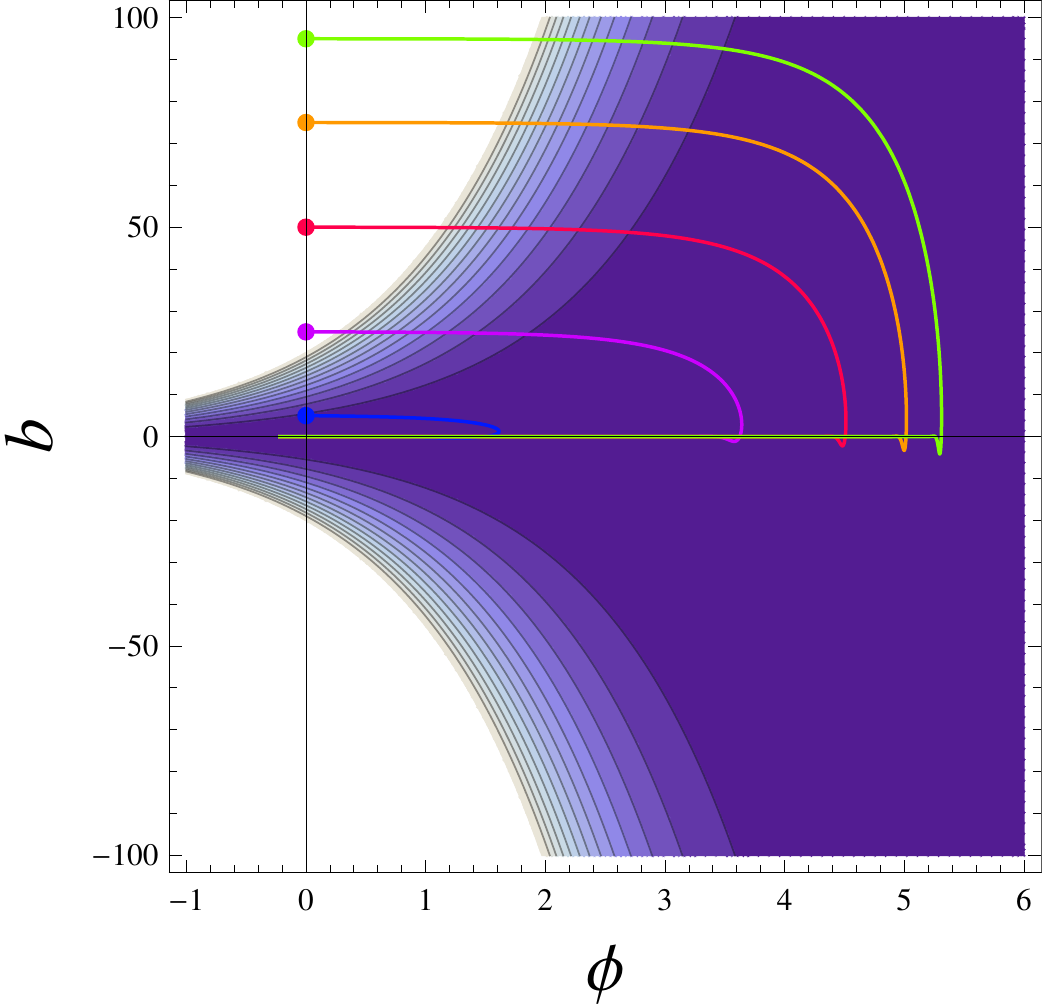}
\caption{\footnotesize 3D plot (left) and contour plot (right) of the potential $V$ in the $(\phi,b)$ plane. Fields are shown in Planckian units, the height of the potential is shown in units $M = 1$. The lines in the contour plot  show the evolution of the fields from its initial values at $\phi = 0$ and various values of $b$. As we see, initially the field $b$ practically does not change, whereas the field $\phi$ rapidly grows. Then the field $b$ rapidly decreases, and the field $\phi$ starts rolling back to the minimum of the potential at $\phi = 0$. This last stage is the only stage where inflation happens; it is described by the usual Starobinsky-Whitt potential, which leads to $r \sim 0.004$.}
\label{fig1}
\end{figure}

One can confirm these tentative and approximate conclusions by solving equations of the fields $\phi$ and $b$ numerically, by extending the framework in \cite{Kallosh:2004rs} where the cosmological  evolution due to the axion-dilaton pair was studied. The results of our calculations for various initial values of the field $b$, are represented in Fig. \ref{fig1}. We present below only examples where $\phi$ starts at its minimum at $\phi=0$, however, we have run the code for various other choices of the initial values of the dilation, we have got the same picture.

A detailed time evolution of each field is shown in Fig. \ref{fig2}. Different colors indicate different initial conditions. We always put $\phi_{\rm initial}=0$ at its minimum value, while $b$ starts from several large values: $b_{\rm init} = 1$ (blue), $b_{\rm init} = 6$ (pink), $b_{\rm init} = 12$ (red), $b_{\rm init} = 30$ (orange), $b_{\rm init} = 48$ (green).

\begin{figure}[h!t!]
\centering
\includegraphics[width=.42\textwidth]{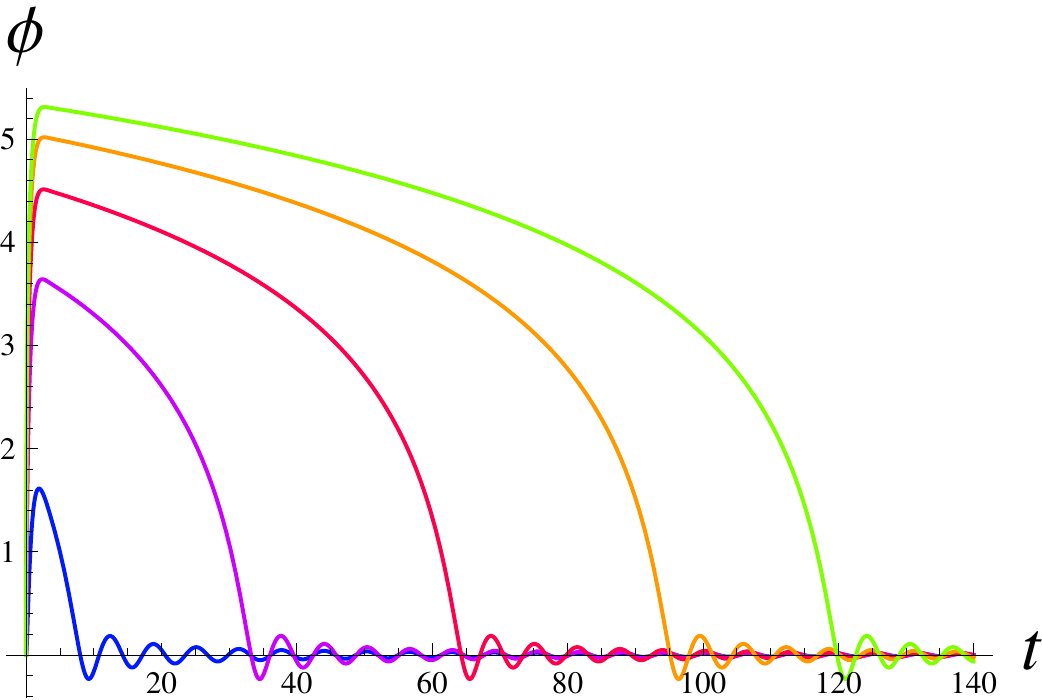}
\hspace{.05\textwidth}
\includegraphics[width=.42\textwidth]{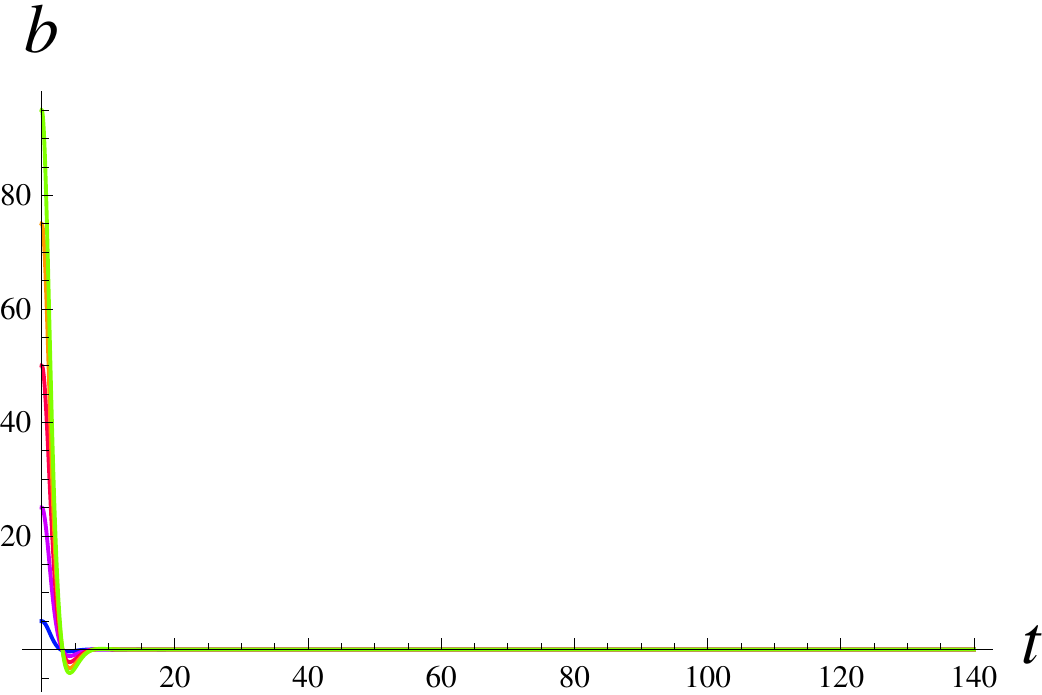}
\caption{\footnotesize Time evolution of the scalar fields, with respect to time measured in units $M^{-1}$.  Immediately, $b$ decreases rapidly, while  $\phi$ increases, showing that it is inconsistent to fix $\phi$ at the minimum. From $t\simeq 3$, the field $b$ is negligible and no longer plays  a role. After this time, the scalar $\phi$ drives inflation while rolling down the potential. Once $\phi$ is small, close to its minimum value $\phi =0$, inflation ends and $\phi$ oscillates in a damped fashion around the minimum.
}
\label{fig2}
\end{figure}

Fig. \ref{fig3} shows expansion of the universe and the behavior of the Hubble constant. As we see, the first stage of rapid expansion of the universe does not lead to an appreciable expansion of the universe, which means that the scenario outlined in \cite{Ferrara:2014ima} does not actually work.  The stage of inflation occurs with a practically constant value of the Hubble constant, just as it should be in the Starobinsky model, where the value of $V$ changes just a little during the main part of the process.
\begin{figure}[h!t!]
\centering
\includegraphics[width=.42\textwidth]{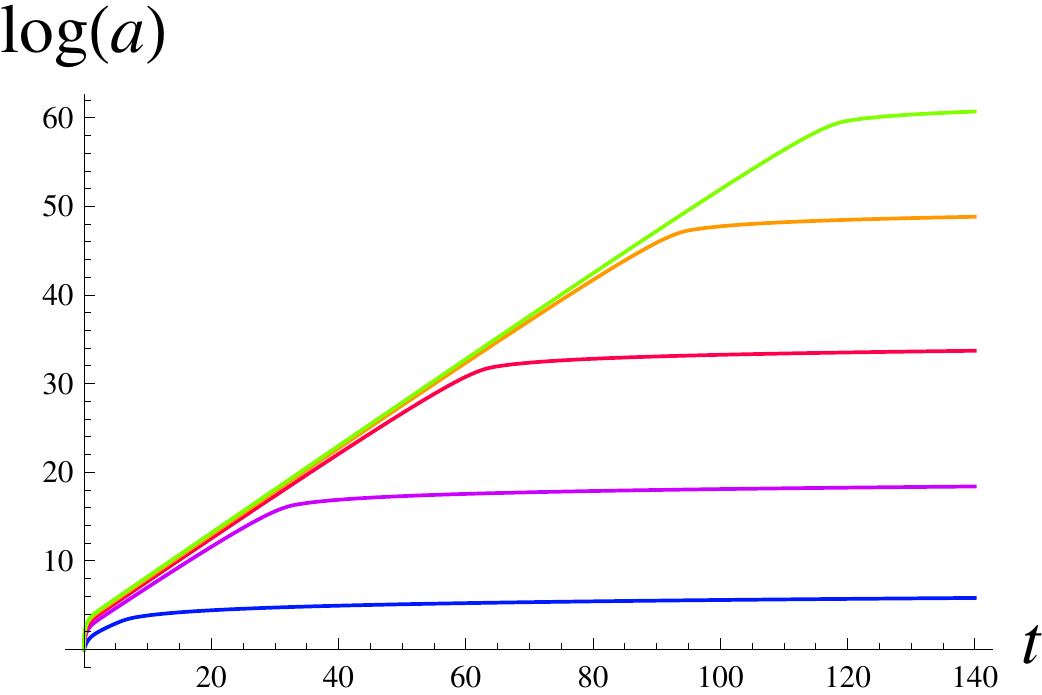}
\hspace{.05\textwidth}
\includegraphics[width=.42\textwidth]{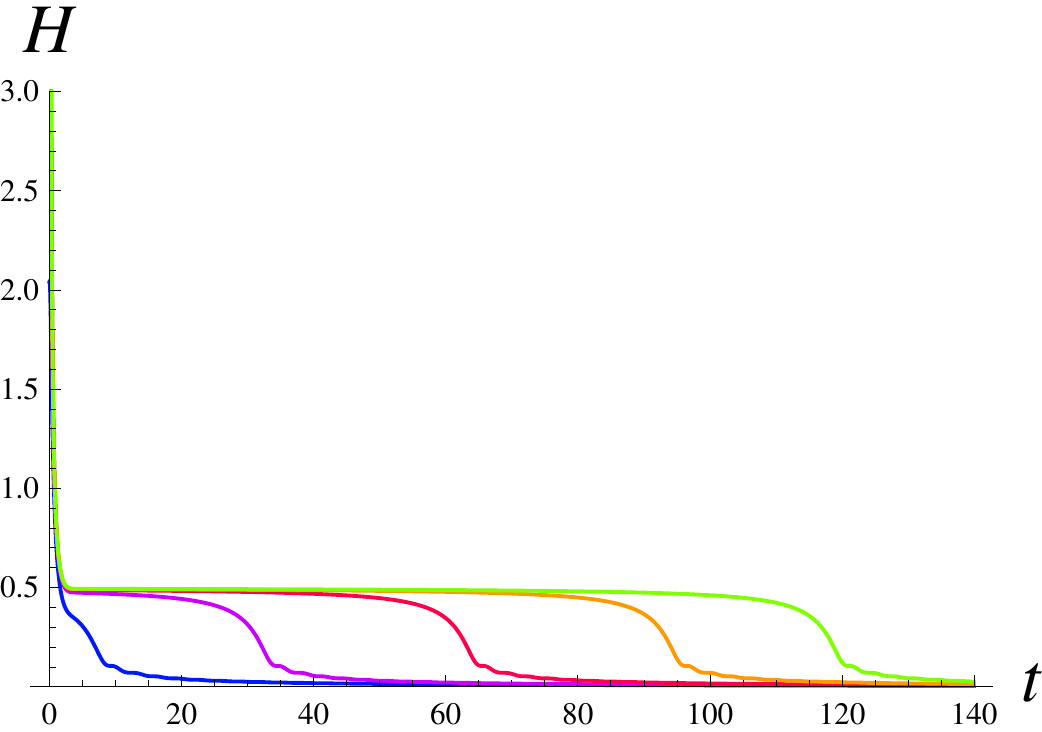}
\caption{\footnotesize  Cosmological evolution  of scale factor (left) and Hubble `constant' (right). After an initial sudden but insignificant increase in $\log a$ (drop in $H$) driven by the field $b$, inflation starts around $t\simeq 3$ with $\phi$ acting as the inflaton, as in the usual Starobinsky model. This is seen by the linear increase of $\log a \sim t$ and a nearly constant $H$.}
\label{fig3}
\end{figure}

\section{Towards chaotic inflation in a modified Cecotti model}

The results obtained above show that the main reason of the absence of chaotic inflation regime in the Cecotti model is the insufficient stabilization of the field ${\rm Re}\ T$ by the Starobinsky-Whitt potential. As we will show now, one can solve this problem and achieve a regime of chaotic inflation with large $r$ by strongly stabilizing the field  ${\rm Re}\ T$, which requires modification of the Starobinsky-Whitt potential. While it is possible to do so, the properly stabilized supersymmetric Cecotti model loses any relation to the Starobinsky model and to modified pure supergravity.

As an example of a model with stabilized ${\rm Re}\ T$, one can consider a model which is obtained from (\ref{K1}) by adding the term $a C\bar C (T+\bar T)$ under the logarithm, and considering a more general superpotential:
\be
\label{K1a}
K = -3\, \ln\Big{(}T + \overline{T} - \overline{C} C + a\, C\bar C\, (T+\bar T)+b\, (C\bar C)^{2}\Big{)}\ , \qquad W= 3 M C f(T-1) \ ,
\ee
where $f$ is some holomorphic function of $T-1$. The last two terms in the \K\ potential are added for stabilizing the fields $t = {\rm Re}\ T$ and $C$. 

One can show that the potential corresponding to this theory at $C = 0$ is given by
\be
V(T) = {3 M^2}\ {|f(T-1)|^{2}\over (T+\bar T)^{2} (1-a(T+\bar T))} \ .
\ee
For example, for the simplest choice $f(T-1) = T-1$ one finds, in terms of the fields  $t = {\rm Re}\ T$ and $s = {\rm Im}\ T$,
\be
V(T) = {3 M^2\over 4}\ {(t-1)^{2}+s^{2}\over t^{2} (1-2at)} \ .
\ee
This potential coincides with the potential of the Cecotti model in the limit $a\to 0$, in terms of the canonically normalized variables, but in that case, as we already checked, inflation happens only during the evolution in the  ${\rm Re}\ T$ direction. With an increase of $a$, the potential in the ${\rm Re}\ T$ direction at large $T$ becomes more steep, and the predictions of the model change. Importantly, there is an infinite potential barrier at $t = 1/(2a)$. With an increase of $a$ towards $a = 0.5$, this barrier comes closer to the minimum of the potential at $t =1$, and the field $t$ becomes strongly stabilized there. The potential of the field $s$ for fixed $t$ remains quadratic, and the simplest regime of chaotic inflation can be realized in this case. For $a < 0.5$, the field $t$ slightly deviates from $t = 1$ during inflation, so the potential along the inflationary trajectory is only approximately quadratic, but the difference from the purely quadratic case disappears in the limit $a \to 0.5$. For a more general choice of the function $f(T-1)$ one can achieve any desirable shape of the chaotic inflation potential.

This particular regime provides yet another generalization of the general mechanism of achieving chaotic inflation in supergravity along the lines of Refs. \cite{Kawasaki:2000yn,Kallosh:2010ug,Kallosh:2010xz}, where inflation occurs in the flat direction of the \K\ potential. In the case considered above, the \K\ potential does not depend on ${\rm Im}\ T$.  The new version of this scenario described above is more complicated than the versions considered before because one is still forced to study a combined evolution of two fields, ${\rm Re}\ T$ and ${\rm Im}\ T$. At any rate, according to \cite{CK}, the resulting theory is no longer related to the theory $R + R^{2}$, even with higher curvatures there are matter multiplets present.

\section{Conclusions}

A unique supersymmetric model with higher curvature (up to 4-derivative terms) without matter was constructed by Cecotti in \cite{Cecotti:1987sa}. It is dual to a standard 2-derivative supergravity interacting with two chiral matter multiplets. This model provides a supersymmetric generalization of the Starobinsky model \cite{Starobinsky:1983zz}, when the imaginary part of the modulus $T$ vanishes and the real part of $T$ evolves \cite{Kallosh:2013lkr}. However this regime produces very low amplitude of tensor modes, which is strongly disfavored by BICEP2.

It was argued in \cite{Ferrara:2014ima} that if one studies an opposite regime, when the imaginary part of $T$ evolves and the real part of $T$ vanishes, the same model   \cite{Cecotti:1987sa} leads to realization of the simplest chaotic inflation scenario \cite{chaotic} with predictions matching the BICEP2 results. This would be a very attractive feature of the model  \cite{Cecotti:1987sa}: It would be able to describe two radically different inflationary regimes, corresponding to two simplest models of inflation,  \cite{chaotic} and \cite{Starobinsky:1983zz}.

However, we have shown here  that the  general regime where both the real and imaginary parts of $T$ are allowed to evolve leads to the same predictions as  the usual Starobinsky model. If the BICEP2 result and their interpretation in \cite{bicep}  are valid, one concludes that the higher derivative pure curvature supergravity \cite{Cecotti:1987sa} does not describe the early universe inflation.  Some matter multiplets have to be present, even in models with higher curvatures, as argued in \cite{CK}, where the relevant chaotic inflation models supporting inflation with high level of gravity waves were studied both in the standard 2-derivative supergravity as well as when higher derivatives terms are  present.

This does not mean that the model  \cite{Cecotti:1987sa} cannot be generalized in such a way as to describe chaotic inflation with a quadratic or nearly quadratic potential. Indeed one can do so e.g. by modifying the \K\ potential of the theory. For example, one can strongly stabilize the real part of the field $T$ by adding to the \K\ potential a term such as $S\bar S (T+\bar T)$ or $S\bar S (T+\bar T)^2$ under the logarithm. We checked that in this case the real part of the field $T$ becomes strongly stabilized at the minimum of its potential, and chaotic inflation driven by the imaginary part of the field $T$ becomes possible. The potential along the inflationary trajectory will be approximately quadratic. However, strong stabilization of the real part of the field $T$ is possible only at the expense of a dramatic modification of the potential. As a result, the stabilized version of the model \cite{Cecotti:1987sa} entirely loses its original relation to the theory $R +R^2$ and the Starobinsky model. This conclusion is fully consistent with the recent results obtained in \cite{CK}.

\

\subsubsection*{Acknowledgments}

We acknowledge stimulating discussions with S. Cecotti,  S. Ferrara, D. Roest,  A. Van Proeyen. RK, AL, BV and WC are supported by the SITP, RK and AL are also supported by the NSF Grant PHY-1316699 and RK is also supported by the Templeton foundation grant `Quantum Gravity Frontiers.' WC is supported by the ArabFund for Economic and Social Development. BV greatfully acknowledges the Fulbright Commision Belgium for financial support. BV is a Fulbright Visiting Scholar and an Honorary Fellow of the Belgian American Educational Foundation.

\end{document}